\documentclass[prl,twocolumn,aps,superscriptaddress,showpacs]{revtex4}

\usepackage{graphicx}
\usepackage{amsmath}
\usepackage{amsfonts}
\begin{document}

\def\be{\begin{equation}}
\def\ba{\begin{eqnarray}}
\def\ee#1{\label{#1}\end{equation}}
\def\ea#1{\label{#1}\end{eqnarray}}
\def\la{\langle \! \langle}
\def\ra{\rangle \! \rangle}
\def\bs{\begin{center}}
\def\es{\end{center}}
\def\fpa#1{\frac{\partial}{\partial #1}}
\def\ve{\varepsilon}
\def\De{D_{eff}}
\def\td{\textrm{d}}
\def\he{\hbar/2e}
\graphicspath{{img/}}
\title{Absolute negative mobility induced by thermal equilibrium fluctuations}

\author{L. Machura}
\affiliation{Institute of Physics, University of Silesia,
40-007 Katowice, Poland}
\author{M. Kostur}
\affiliation{Institute of Physics, University of Augsburg,
D-86135 Augsburg, Germany}
\author{P. Talkner}
\affiliation{Institute of Physics, University of Augsburg,
D-86135 Augsburg, Germany}
\author{J. \L uczka}
\affiliation{Institute of Physics,  University of Silesia,
40-007 Katowice, Poland}
\author{P. H\"anggi}
\affiliation{Institute of Physics, University of Augsburg,
D-86135 Augsburg, Germany}
\begin{abstract}
  A novel transport phenomenon is identified that is induced by
  inertial Brownian particles which move in simple one-dimensional,
  symmetric periodic potentials under the influence of both a time
  periodic and a constant, biasing driving force.  Within tailored
  parameter regimes, thermal equilibrium fluctuations induce the
  phenomenon of absolute negative mobility (ANM), which means that the
  particle noisily moves {\it backwards} against a small constant
  bias. When no thermal fluctuations act, the transport vanishes
  identically in these tailored regimes. There also exist parameter
  regimes, where ANM can occur in absence of fluctuations on grounds
  which are rooted solely in the complex, inertial deterministic
  dynamics. The experimental verification of this new transport scheme
  is elucidated for the archetype symmetric physical system: a
  convenient setup consisting of a resistively and capacitively
  shunted Josephson junction device.
\end{abstract}
\maketitle

A central result of thermodynamics is due to
Henri Louis Le Chatelier which (loosely speaking) states that "a
change in one of the variables that describe the system at
equilibrium produces a shift in the position of the equilibrium that
counteracts this change". In particular, if a system is at thermal
equilibrium, its reaction to an applied bias is so, that the
response is  in the same direction of this applied force, towards a
new equilibrium.

Thus, the seemingly paradoxical situation that the system's response
is opposite to a small external load is prohibited by the laws of
thermodynamics; it would imply the phenomenon of an absolute
negative mobility (ANM). A possibility to circumvent the stringent
conditions imposed by thermodynamics is, however, to go far from
equilibrium where these restrictions no longer possess validity.
Known examples of such absolute negative mobilities or likewise,
absolute negative conductivity, have been experimentally observed
before within a quantum mechanical setting in p-modulation-doped
multiple quantum-well structures \cite{hop}, in semiconductor
superlattices \cite{keay} and have been studied as well
theoretically for ac-dc-driven tunnelling transport
\cite{HarGri1997}, in the dynamics of cooperative Brownian motors
\cite{BroBen2000}, for Brownian transport containing a complex
topology \cite{EicRei2002a,EicRei2002b} and in some stylized,
multi-state models with state-dependent noise \cite{CleBro2002}, to
name but a few.

This startling and counterintuitive transport phenomenon has spurred
renewed interest, motivated by the quest to explore new transport
scenarios in simple and easy to fabricate devices that intrinsically
exploit the abundant source of thermal fluctuations to a
constructive technological use. Physical systems that are most
suitable for this purpose are periodic one-dimensional symmetric
systems such as phase differences across Josephson junctions
\cite{junction}, rotating dipoles in external fields \cite{Reg2000,
Coffey}, superionic conductors \cite{Ful1975} and charge density
waves \cite{Gru1981}. Another important area constitutes the
noise-assisted transport of Brownian particles \cite{RevMod,
ChaosSI}, as it occurs for Brownian motors possessing ample
applications in physics and chemistry \cite{BM}.

Our main objective is to detect noise-induced, absolute negative
mobility (ANM) in ordinary physical systems that are readily available
and which can be put to immediate use without the need to go to
extremely low temperatures and /or the use of advanced fabrication
techniques of higher-dimensional stylized structures that generate the
necessary trapping mechanism for the occurrence of ANM, cf. the nicely
tailored two-dimensional trap-geometries used in Refs.
\cite{EicRei2002a,EicRei2002b}. Our minimal prerequisites for
detecting ANM therefore are: the use of (i) simple, one-dimensional
symmetric periodicity, (ii) symmetric external forcing, (iii) inertial
dynamics and (iv) thermal fluctuations.  The following model set-up
fulfills all of these.

We formulate the problem in terms of a Brownian classical particle
of mass $M$ moving in a spatially periodic potential $V(x)=V(x+L)$
of period $L$ and barrier height $\Delta V$, subjected to an
external, {\it unbiased} time-periodic force $A \cos (\Omega t)$
with angular frequency $\Omega$ and of amplitude strength $A$.
Additionally, a constant external force $F$ acts on the system.

The dynamics of the system is thus modeled by the inertial Langevin
equation \cite{HanTho1982}
\begin{eqnarray} \label{Lan1}
M \ddot x + \Gamma \dot x = -V'(x) + A \cos(\Omega t) +F
+ \sqrt{2\Gamma k_B T} \; \xi(t),
\end{eqnarray}
where a dot and prime denote differentiation with respect to time $t$
and the Brownian particle's coordinate $x$, respectively.  The
parameter $\Gamma$ denotes the friction coefficient, $k_B$ the
Boltzmann constant and $T$ is the temperature. Thermal fluctuations
due to the coupling of the particle with the environment are modeled
by zero-mean, Gaussian white noise $\xi(t)$ with auto-correlation
function $\langle \xi(t)\xi(s)\rangle = \delta(t-s)$. The spatially
periodic potential $V(x)$ is assumed to be {\it symmetric} and chosen
in its simplest form, namely,
\begin{eqnarray}
\label{pot}
V(x) =  \Delta V \sin (2\pi x/L).
\end{eqnarray}
Upon introducing the period $L$ and the parameter combination $\tau_0=
L \sqrt{M/\Delta V}$ as units of length and time \cite{MachuraJPC},
respectively, Eq. (\ref{Lan1}) can be rewritten in dimensionless form,
reading
\begin{equation}
\label{Lan2}
\ddot{\hat{x}} + {\gamma} \dot{\hat x} =- \hat{V}'(\hat{x}) + a
\cos(\omega \hat{t}) + f +\sqrt{2{\gamma}D_0} \; \hat{\xi}
(\hat{t}),
\end{equation}
where $\hat{x} = x/L$ and $\hat{t} = t/\tau_0$.  The remaining
re-scaled parameters are: the friction coefficient ${\gamma} = (\Gamma
/ M) \tau_0$, the potential $\hat{V}(\hat{x})=V(L\hat{x})/\Delta V =
\hat{V}(\hat{x}+1) = \sin (2\pi\hat{x})$ with unit period and barrier
height $\Delta \hat{V}=2$, the amplitude $a = L A / \Delta V$, the
frequency $\omega = \Omega \tau_0$, the load $f=L F/ \Delta V$, the
zero-mean Gaussian white noise $\hat{\xi}(\hat{t})$ with
auto-correlation function
$\langle\hat{\xi}(\hat{t})\hat{\xi}(\hat{s})\rangle=\delta(\hat{t}-\hat{s})$,
and the noise intensity $D_0 = k_B T / \Delta V$.  From now on, we
will only use the dimensionless variables and shall omit the ``hat''
for all quantities occurring in Eq.  (\ref{Lan2}).

\begin{figure}[htpb]
  \begin{center}
    \includegraphics[angle=0,width=0.99\linewidth]{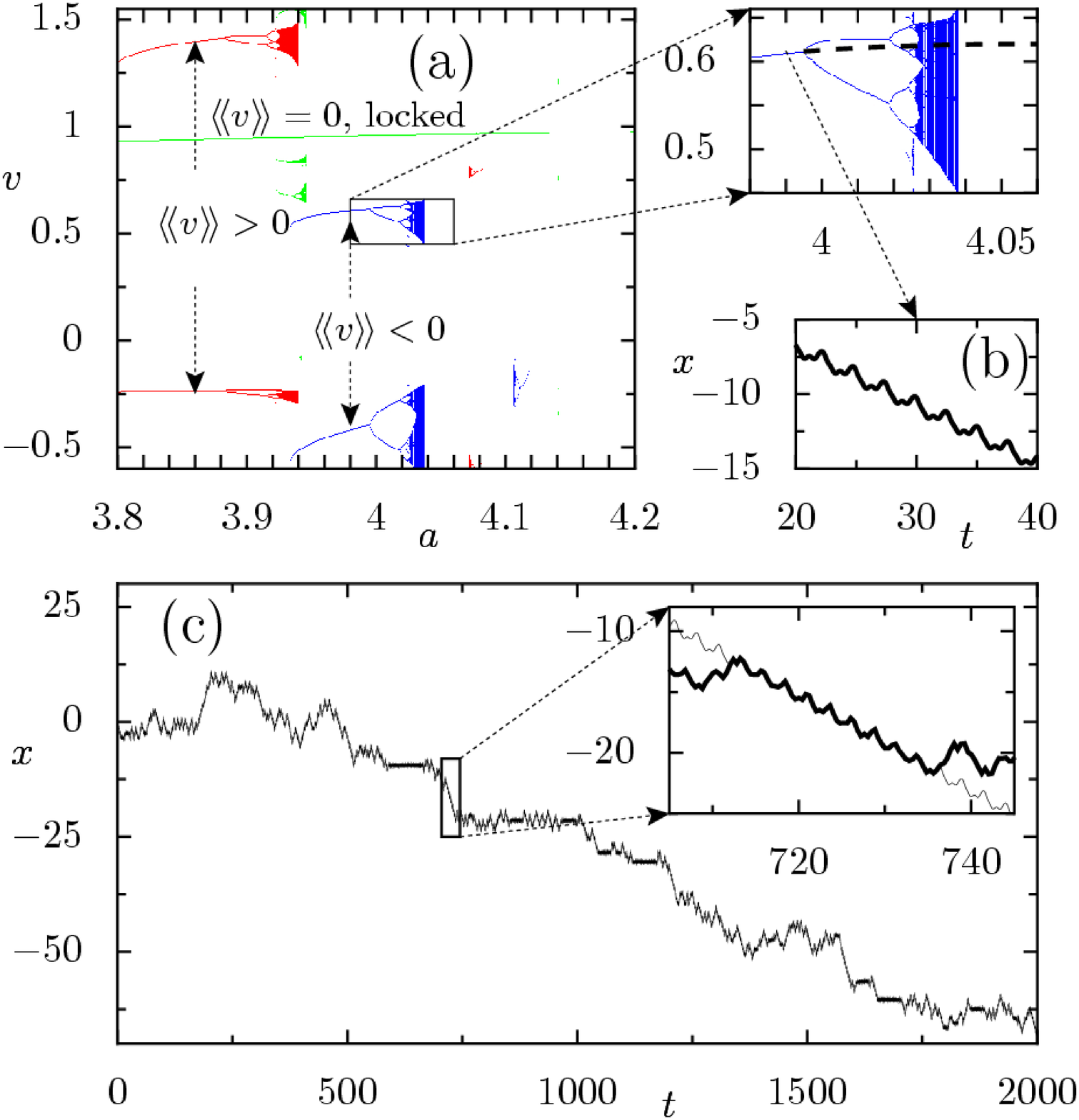}
  \end{center}
  \caption{(Color online) In panel (a) we present a small part of the
    bifurcation diagram of the noiseless system as a function of the
    amplitude $a\in(3.8,4.2)$. The ordinate indicates the stroboscopic
    velocity in the asymptotic long time limit. The system is biased
    by the positive force $f=0.1$. The other parameter values are
    $\gamma=0.9$ and $\omega =4.9$. The straight line at $v \approx
    0.9$ corresponds to a stable locked state. It persists for all
    shown values of the driving amplitude.  For amplitudes up to
    $a\approx 4.04$ a running state with negative average velocity
    coexists. In panel (b) the corresponding orbit is depicted for
    $a=3.99$. It is a period 2 state which stays within one potential
    well in the first period and moves to the left neighboring state
    in the second period.  In panel (c) a realization of the
    stochastic system is displayed for $a=4.2$ and $D_0=0.001$. Its
    average velocity is negative. In the inset we show a typical part
    of the stochastic trajectory (dotted line) which contributes to
    the transport. For several periods of the driving force it closely
    follows a deterministic unstable periodic orbit (solid line). This
    orbit lies on the unstable branch emanating from a pitchfork
    bifurcation at $a\approx 3.99$ of the running state shown in panel
    (b). A part of the bifurcation diagram including the first
    bifurcation of the period two orbit to a stable period four orbit
    (solid line) and the mentioned unstable period two orbit (dotted
    line) is shown in the magnification in panel (a). }
  \label{fig1}
\end{figure}

This Langevin equation (\ref{Lan2}) provides a simple model of the
diverse periodic systems specified in the introduction.  Yet the
deterministic inertial dynamics of the nonequilibrium system defined
by Eq. (\ref{Lan2}) exhibits a very rich and complex behavior
\cite{kautz,jung96,Mateos}. Depending on the parameter values,
periodic, quasiperiodic and chaotic motion can result in the
asymptotic long time limit, see Fig.~\ref{fig1}. Different initial
conditions of position and velocity can also lead to different
asymptotic behavior, i.e. various attractors may coexist.

A rough classification of the asymptotic behavior can be made into
locked states in which the motion is confined to a finite number of
spatial periods, and running states in which the motion is unbounded
in space. For the deterministic transport properties the running
states are crucial. A broad spectrum of various forms of running
states exists which comprises -- apart from periodic -- also chaotic
motions.  By adding thermal fluctuations, one typically activates a
diffusive dynamics leading to random transitions between possibly
coexisting basins of attraction, which play an analogous role to
potential wells in equilibrium systems. For example, stable locked
states of the deterministic system are destabilized by noise:
transitions between neighboring locked states will lead to diffusive
or even directed transport.

The most important quantifier for characterizing directed transport is
the asymptotic mean velocity $\la v \ra$ \cite{MachuraJPC} which is
defined as average of the velocity over the time and thermal
fluctuations.  The Fokker-Planck equation corresponding to Eq.
(\ref{Lan2}) cannot be analytically solved, therefore we carried out
extensive numerical simulations of the Langevin equation. Details of
the employed numerical scheme are described in
Ref.~\cite{machurathesis}.
Parts of our so obtained results are presented
next.

The velocity of a stable running state mostly points into the
direction of the force $f$. But there are also running stable states
which on average move in the {\it opposite} direction of the
constant driving force, hence displaying deterministic ANM
\cite{ReimannNote}, see Fig.~\ref{fig1}(b). In these deterministic
cases the energy consumed for moving uphill is taken from the
oscillating driving force.

\begin{figure}[htpb]
  \begin{center}
    \includegraphics[angle=0,width=0.9\linewidth]{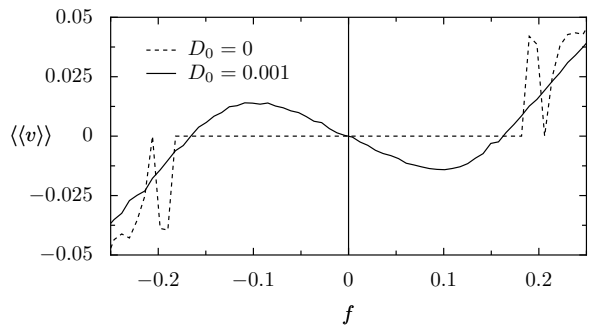}
  \end{center}
  \caption{The average velocity $\la v \ra$ of an inertial Brownian
    particle described by eq. (\ref{Lan2}) is depicted as a function
    of the external force $f$ for the deterministic (dashed line) and
    noisy (solid line) dynamics. The system parameters are: $a=4.2$,
    $\omega=4.9$, $\gamma=0.9$ and $D_0=0$ (dashed line) and
    $D_0=0.001$ (solid line) The absolute mobility defined as $\la v
    \ra/ f$ assumes a negative value for the noisy system in the range
    $|f|<0.17$.  The most pronounced ANM occurs for small absolute
    values of the bias $f$.  For the bias force $f \in
    (-f_{\text{stall}},f_{\text{stall}})$ with $f_{stall} \approx \pm
    0.17$, the Brownian particle moves opposite to the applied bias $f$. }
  \label{fig2}
\end{figure}

Most remarkably, within particular parameter regimes ANM is solely
induced by thermal noise. In the case depicted in Fig.~\ref{fig2} the
deterministic average velocity vanishes for a whole range of forces
around zero, whereas a very small amount of noise yields a negative
mobility at small forces. The region of ANM is bounded by the stall
forces $\pm f_{\text{stall}}$ for which the average velocities vanish.
With increasing temperature the stall force decreases and wears off at
a finite temperature.

Although the above described ANM only manifests itself in presence of
thermal fluctuations, the underlying relevant mechanisms are strongly
influenced by the deterministic dynamics of the system.  At the
driving strength $a=4.2$ the deterministic system only possesses one
stable orbit with period one which is a locked state and therefore
does not contribute to the transport. Consequently, the current at
$D_0=0$ is zero. There exists, however, a large number of unstable
periodic orbits, transporting the particle in both positive and
negative directions, which influence the relaxation dynamics from
points lying far form the stable locked orbit. In presence of noise,
the particle is permanently moved away from the stable orbit. In Fig.
\ref{fig1}(c) we depict a single realization of the stochastic
dynamics at $D_0=0.001$. One observes a systematic movement of the
particle position into the negative direction. Moreover, between noisy
bursts, periods of almost regular motion take place. All these regular
parts are distinguished by approximately the same negative average
velocity and therefore primarily contribute to the negative mobility.
One of these regions has been blown up in the inset in Fig.
\ref{fig1}(c).  The random trajectory strongly resembles a particular
unstable orbit of the deterministic system. Upon reducing the driving
amplitude $a$ this orbit can be identified as the unstable branch
emerging from a pitchfork bifurcation of a stable orbit, see the inset
of Fig.~\ref{fig1}(a). Before the bifurcation, the corresponding
stable orbit also has negative average velocity. It is depicted in
Fig.~\ref{fig1}(b). Among many other unstable periodic orbits this
unstable remnant of a stable running state apparently is most likely
populated by the noise.

\begin{figure}[htpb]
  \begin{center}
    \includegraphics[angle=0,width=0.9\linewidth]{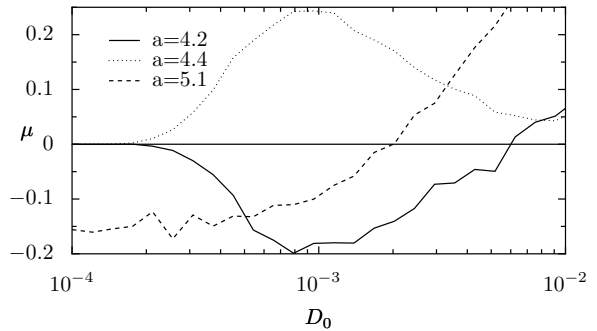}
  \end{center}
  \caption{The mobility coefficient
$\mu = (\partial \la v \ra/ \partial f)(f=0)$
 depicted versus the dimensionless
  temperature strength, $D_0 \propto T$,
 for three values of the cosine-driving strength $a$; the strength
 $a  = 4.2$ (solid) corresponds to thermal noise induced ANM. The driving strength $a=
 5.1$ (dashed) corresponds to a regime exhibiting deterministic ANM and $a= 4.4$
 (dotted) is for a normal nonlinear response regime.
   The remaining parameters
  are: $\omega=4.9$, $\gamma=0.9$.   }
  \label{fig3}
\end{figure}

In Fig. 3, we depict the dependence of the mobility coefficient $\mu =
(\partial \la v \ra/ \partial f)(f=0)$ versus temperature for three
cases: (i) the noise-induced ANM ($a= 4.2$), (ii) a regime with
deterministic ANM ($a=5.1$) and (iii) the 'normal' or 'positive'
mobility regime ($a=4.4$), when the velocity is positive for positive
load, i.e. $\mu > 0$. For $a=4.2$, there emerges an optimal
temperature at which ANM is the most prominent. For $a=5.1$, a
continued increase of temperature eventually annihilates ANM. For the
amplitude strength $a=4.4$, an optimal temperature occurs at which the
mobility is maximal. There also occur regimes of so termed {\it
  differential} negative mobility \cite{machurathesis}; however, a
more complete analysis is beyond this letter format but will be
presented elsewhere. In particular, the effect of noise-induced ANM in
(\ref{Lan1}) does not present an exception but can emerge as well in
different regimes.

As an application of the above theoretical study we consider  the
resistively and capacitively shunted single Josephson junction for
which the Absolute Negative Conductance (ANC) can be measured, thus
putting our predictions to a reality check.  The phase difference
$\phi$ across the junction obeys Eq.~(\ref{Lan1}) with $x=\phi
-\pi/2$, the mass is $M=(\he)^2 C$, the friction coefficient $\Gamma =
(\he)^2(1/R)$, the barrier hight $\Delta V = E_J=(\he) I_0$ and the
period $L=2\pi$, where $C$ denotes the capacitance, $R$ the
normal-state resistance of the junction, $E_J=(\hbar/2e)I_0$ the
coupling energy of the junction and $I_0$ the critical current. The
load $F=(\he) I_{d}$ corresponds to a dc-bias current, whereas the
amplitude strength $A=(\he) I_{a}$ and the frequency $\Omega$ define
the external ac-"rocking" current.  The average velocity $\la v \ra$
translates into the voltage $\mathbb{V}=(\hbar/2e)\omega_0 \,\la v \ra
$ across the junction, wherein $\omega_0=(2/\hbar)\sqrt{E_JE_C}$ is
the Josephson plasma frequency with the charging energy $E_C=e^2/C$.
The dimensionless noise intensity $D_0 = k_B T/ E_J$. Given the above
relation, the dc-current-voltage characteristics can be obtained in
the same form as presented in Fig. 2. The experimental setup to
observe our novel noise induced ANC necessitates the following
parameter values: the amplitude of the ac-current $I_a = aI_0/2\pi
\approx 0.67 \;I_0$, an angular driving frequency $\Omega =\omega
\omega_0/2\pi \approx 0.78\;\omega_0$, a bias $I_d = f I_0/2\pi$ which
becomes $I_d \approx 0.016 \; I_0$ for the bias $f \approx 0.1$ which
leads to an extremal velocity $\la v \ra$, see Fig.~\ref{fig2}. The
frequency $\omega_r=1/\tau_r=1/RC$ which gives the relaxation time
$\tau_r$ is $\omega_r =(\gamma/2\pi) \;\omega_0 \approx 0.14
\;\omega_0$ and temperatures is of the order $k_B T = E_J D_0 \approx
0.001\; (\hbar/2e)I_0$. For example, for a junction possessing a
critical current of $I_0 = 0.1$mA, a resistance $R=2.9 \Omega$ and a
capacitance of $C=20$pF, the dc-bias becomes $I_d=1.6 \mu$A, the
amplitude of the ac-current is $I_a=67 \mu$A, and the driving
frequency is in the GHz-regime, i.e. $\Omega = 96$GHz. The optimal
temperature occurs at $T=2.4$K and the maximal voltage is
$\mathbb{V}=0.6\mu$V.  In turn, we predict a maximal voltage
$\mathbb{V}=2\mu$V for a junction at the temperature $T=24$K with $I_0
= 1$mA, $R=1 \Omega$ and $C=16$pF, a dc bias $I_d=16 \mu$A, and an ac
current with amplitude $I_a=0.67$mA and frequency $\Omega = 340$GHz.

Our work has demonstrated that the surprising effect of a solely
thermal {\it noise-induced} absolute negative mobility can occur in
generic, biased {\it symmetric} systems and devices that can be
described by Eq. (\ref{Lan1}). In clear contrast, for non-inertial,
overdamped systems (i.e. with $M\ddot x =0$) ANM cannot occur; this
fact underpins the crucial role that inertial effects play for this
anomalous transport feature. Moreover, the presence of a
non-adiabatic, high-frequency external driving is an indispensable
ingredient for ANM. Notably, the phenomenon of ANM implies with $ F
\la v\ra < 0$ that beneficial power can be extracted upon the
switch-on of a positive load force $F$. This phenomenon of either
pure noise-induced ANM or deterministic ANM must clearly be
distinguished from the phenomenon of a noise-assisted, directed
transport occurring in Brownian motors \cite{BM} which use a
symmetry-breaking of either spatial or temporal origin; for
(uncoupled) Brownian motors we have that for a bias $F=0$ the
current is finite (with a positive mobility) while for ANM it is
zero (with negative mobility).

These surprising ANM-findings can readily be experimentally tested
with a single Josephson junction device when driven in the
experimentally available GHz-regimes. Other test systems described by
the model (\ref{Lan1}) are superionic conductors or
rotating dipoles in external fields.\\

Work supported by the Deutsche Forschungsgemeinschaft (DFG) via
grant HA 1517/13-4, the DFG-SFB 486, ESF and  PBZ-MIN-008/P03/2003
(MNiI, Poland).

\end{document}